# Design of GNSS-RTK Landslide Monitoring System Based on Improved Raida Criterion

Junming Wang[a], Yi Shi[b]

[a]School of Computer Science, The University of Hong Kong, Hong Kong SAR, China;
[b]School of Electronics and Information Engineering, Beijing Jiaotong University, Beijing, China, 100091

## ABSTRACT

Aiming at the problem that GNSS-RTK technology cannot effectively monitor landslides due to gross errors and high-frequency noise during landslide monitoring, a GNSS-RTK landslide monitoring system based on the improved Raida criterion(3σ) was designed. The system uses Raspberry Pi as the control core, GNSS-RTK technology to complete deformation data collection, and combines NB-IoT to achieve data transmission and cloud storage. To further improve the monitoring accuracy, real-time gross error detection and high-frequency noise removal method based on the improved Raida criterion(3σ) and Butterworth low-pass filtering is proposed, combined with edge computing devices to complete real-time data processing, and reduce the pressure of cloud computing. The experimental results show that the system's data transmission is reliable and efficient. After gross error elimination and noise processing, real-time deformation monitoring of 10 mm in the horizontal and vertical directions is achieved; simultaneously, the monitoring error of the GNSS combined system is smaller than that of the Single-GPS system.

## 1. INTRODUCTION

Landslides are one of the largest and most frequent disasters among China's geological disasters. The root cause of such disasters is the displacement and deformation inside the soil. There are many ways to monitor landslides. The different monitoring contents are divided into surface deformation and underground deformation monitoring, influencing factors monitoring, macro-geological monitoring, etc. [1]. The GNSS-RTK monitoring method[2] has the advantages of high monitoring accuracy and flexible point selection, which can carry out all-weather and automatic monitoring of small displacements in the earthwork.

Wide range of applications. Huang Guanwen[3]et al. provided a solution for low-cost landslide monitoring with a thousand-yuan terminal, positioning software supplemented by high-efficiency cloud computing. Bai Zhengwei[4] and others developed a thousand-yuan-level miniaturized real-time Beidou/GNSS monitoring device to achieve millimeter-level deformation monitoring. Jessica Glabsch[5] and others combined wireless sensor networks and GNSS to develop a low-cost landslide monitoring system, applied it to actual engineering, and achieved good results. The above studies primarily improve the monitoring methods, R&D costs, and data transmission and storage but still lack data processing and monitoring accuracy. Therefore, this paper proposes a GNSS-RTK landslide monitoring system based on the improved Raida criterion.

## 2. DATA PROCESSING MODEL ESTABLISHMENT

### 2.1 GNSS-RTK principle

GNSS-RTK technology is based on relative positioning to process the carrier phase values of two stations in real-time. That is, one station is placed at a known point as a reference station, and another or several stations are placed at a point to collect data as a monitoring station for simultaneous observation. The carrier phase value received by the reference station and its coordinates are combined through the data link. The information is sent to the mobile station for differential processing, and the coordinates of the observation point are calculated[6-7]. At present, the common difference processing methods are single, double, triple, and so on. In this system, the receiver uses a double-difference model for baseline calculation, in which the carrier phase observation equation of satellite j and station i[8] is as follows:

$$\Phi_i^j(t) = \frac{f_0}{c}\rho_i^j(t) + f_0[\delta t_i(t) - \delta t^j(t)] - N_i^j(t_0) + \frac{f_0}{c}[\Delta_{i,I_p}^j(t) + \Delta_{i,T}^j(t)] \tag{1}$$

In the above formula: $i$ represents the observation station, $j$ represents the satellite, $c$ is the propagation speed of the radio wave, $\delta t_i(t)$ represents the clock error of the observation station, $\delta t^j(t)$ represents the satellite clock error.

From the formula (1), the single-difference observation equation can be obtained as:

$$\Delta\Phi_{21}^j(t) = \frac{f_0}{c}[\rho_2^j(t) - \rho_1^j(t)] + f_0\Delta t_{21}(t) - \Delta N_{21}^j(t_0) \tag{2}$$

The double-difference model is based on the single-difference model to make another difference, eliminating the carrier phase error and the satellite clock error[8]. From the formula(2), the double-difference observation equation can be obtained as:

$$\nabla\Delta\Phi_{21}^{kj}(t) = \frac{f_0}{c}[\rho_2^k(t) - \rho_1^k(t) - (\rho_2^j(t) - \rho_1^j(t))] - \nabla\Delta N_{21}^{kj} \tag{3}$$

## 2.2 Gross Error Removal—Improving the Raida Criterion(3σ)

Gross error detection and elimination are the research hotspots in GNSS deformation monitoring. When eliminating multi-path errors and cycle slip errors in GNSS-RTK monitoring, the Raida criterion is most often used, that is $3\sigma$, the criterion[9], which is mainly used to process monitoring data with large sample size and random errors. The basic steps are:

①Calculate the arithmetic mean of the measured data：

$$\bar{x} = \sum_{i=1}^{n} t_i \tag{4}$$

②Calculate the residual:

$$V_i = X_i - \bar{x}(i = 1,2,3...n) \tag{5}$$

③Calculate the standard deviation:

$$\sigma = \sqrt{\frac{\sum_{i=1}^{n}(x_i - \bar{x})^2}{n-1}} = \sqrt{\frac{(x_1 - \bar{x})^2 + (x_2 - \bar{x})^2 + ... + (x_n - \bar{x})^2}{n-1}} \tag{6}$$

If the residual error of the measured value satisfies: $|V_b| = |X_b - \bar{x}| > 3\sigma$, it is considered $X_b$ as an error value with gross errors and should be eliminated.

Although the Raida criterion has high applicability in gross error detection, there will be a "misjudgment" phenomenon in real-time monitoring, that is, removing deformation data as gross errors will impact real-time deformation monitoring. Therefore, this paper proposes the "time slice" Raida criterion after its improvement. That is, the data in a certain time slice T is processed to solve its confidence interval. If no deformation occurs, the data beyond the confidence interval can be eliminated as gross errors. If deformation occurs, the confidence interval needs to be recalculated to avoid the deformation data being eliminated as gross errors. At the same time, judging whether the data is a gross error or deformed data is determined by the threshold range W derived from expert knowledge. The specific process is shown in Figure 1.

When using the improved Laida criterion on the edge side, whenever a gross error is eliminated, the average, residual, and standard deviation of the data in a "time slice" need to be recalculated, which undoubtedly increases the amount of calculation on the edge side, so this article based on improving the Laida criterion, the monitoring data $t_i$ is used to calculate the standard deviation. The average value $t_0$ is slightly corrected with a fixed amount close to the average value $\bar{x}$ [9-10] in this method:

$$\chi_i = t_i - t_0 \tag{7}$$

average value:

$$\bar{x} = t_0 + \frac{1}{n}\sum_{i=1}^{n}\chi_i \tag{8}$$

Residual:

$$|V_i| = t_i - \bar{x} \tag{9}$$

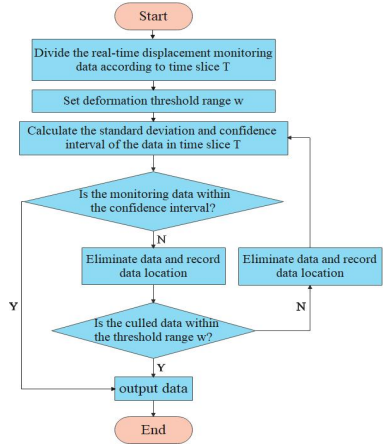

Figure 1. Improve the implementation process of $3\sigma$ criterion

The standard deviation is as follows:

$$\sigma = \sqrt{\frac{1}{n-1}\left[\sum_{i=1}^{n}\chi_i^2 - \frac{1}{n}\left(\sum_{i=1}^{n}\chi_i\right)^2\right]} \quad (10)$$

It is not difficult to find that when using the above four equations to eliminate gross errors, $\chi_i$ and $\chi_i^2$ can be calculated simultaneously. After eliminating a gross error, There is no need to recalculate the accumulated value of the original data, thus saving the time for calculating the accumulated sum of the data in the time slice.

### 2.3 Noise reduction processing-Butterworth filter

Low-pass filtering is a signal filtering method that allows low-frequency signals to pass while blocking and spacing high-frequency signals with a set threshold. Due to the presence of high-frequency noise in the GNSS monitoring sequence, low-pass filtering is used to remove high-frequency noise after gross error detection and elimination. In this paper, a Butterworth low-pass filter[12-13] is used to remove high-frequency noise. Its mathematical expression and The modulus square function is as follows:

$$|H(w)|^2 = \frac{1}{1+\left(\dfrac{w}{w_c}\right)^{2n}} \quad (11)$$

$$|H(j\Omega)|^2 = \frac{1}{1+(j\Omega/j\Omega_c)^{2N}} \quad (12)$$

$$|H(j\Omega)|^2 = \frac{1}{1+(\Omega)^{2N}} \quad (13)$$

This method needs to determine the order according to the specified technical indicators[14]:

$$N \geq \frac{\lg\sqrt{(10^{0.1\alpha_s}-1)/(10^{0.1\alpha_p}-1)}}{\lg(\Omega_s/\Omega_p)} \quad (14)$$

The cut-off frequency of the NovAtel OEM628E receiver output is 5Hz, so the value of the order N is 5, that is, the 5th order Butterworth filter is used to process the monitoring data on the edge side.

## 3  SYSTEM DESIGN

### 3.1  Overall design

The overall architecture of the monitoring system is shown in Figure 2. The system consists of three parts: data acquisition layer, data transmission layer, and application service layer.

### 3.2  Implementation of the data collection layer

The data acquisition layer uses RTK technology to construct monitoring stations and reference stations. The main processors of both stations use Raspberry Pi Zero. Its CPU is ARM11, and its memory is 512 MB. Compared with other versions of Raspberry Pi, it has lower power consumption. To facilitate embedded development. The receiver used in this system's monitoring station and reference station is the NovAtel OEM628E receiver, which uses USB serial communication with the Raspberry Pi to read the positioning data and displacement data after RTK calculation. On the edge side, the system uses Raspberry Pi 4B as the edge computing node for gross error detection and noise removal, and the SX1268-LoRa module is used for data transmission between the monitoring station and the edge node.

### 3.3 Data transmission layer realization

In order to store the data on the edge side to the cloud server, the data transmission layer is composed of NB-IoT technology[15-16] and MQTT protocol. The data is uploaded to the Alibaba cloud platform through NB-IoT and stored in the MySQL database. NB-IoT adopts the extended version of Micro Snow Electronics SIM7020C, which supports TCP, UDP, HTTP, MQTT protocols. The workflow of the data transmission layer is shown in Fig. 3.

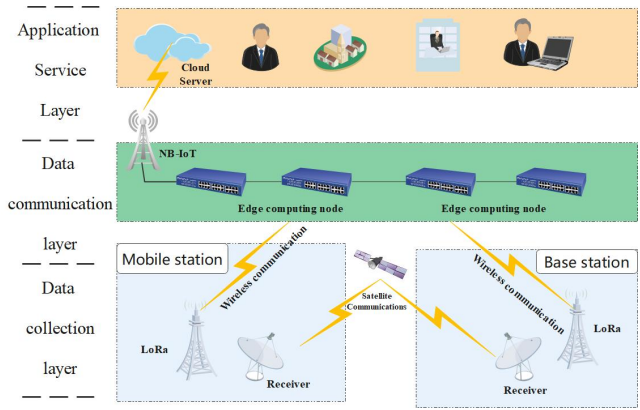
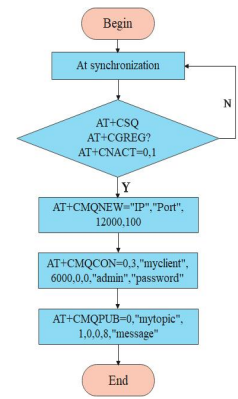

Figure 2. Monitoring system architecture diagram     Figure 3: Data communication flow chart

### 3.4 Application service layer implementation

In order to facilitate users to manage and evaluate the monitoring area remotely, a web terminal is developed based on the PHP language for users to use, which mainly includes functions such as login, registration, user management, authority management, data visualization, and displacement warning. In order to facilitate users to view real-time displacement changes in the monitoring area in real-time, a three-level early warning mechanism is designed for level and elevation direction. When the real-time displacement exceeds the threshold, an early warning will be issued on the Web side.

## 4. EXPERIMENTAL TEST AND ANALYSIS

### 4.1 Experimental plan design

In order to verify the timeliness and reliability of the system in monitoring deformation, a deformation experiment platform was built on the roof of the School of Electronics and Information Engineering of Lanzhou Jiaotong University for testing (Figure 6). The base station and mobile station are encapsulated in an acrylic box, and the satellite antenna is placed outside the acrylic box and placed on the test bench in a relatively short baseline mode. The base station and the mobile station are 800mm apart, and solar panels and batteries are used to provide a stable power supply.

### 4.2 Analysis of comparison results of positioning system performance indicators

Through the statistical analysis of the number of satellites and the change of the factor of precision (DOP) value of the GNSS combined system and the GPS single system in the same epoch during the experimental period, it can be seen that the number of satellites used by the GNSS combined system is kept at about 28. The DOP value is kept below 1; the number of satellites used in the GPS system is kept at about 14, and the DOP value fluctuates at above 0.8. The experimental results show that the DOP value of the GNSS combined system is smaller than that of the Single-GPS system, and it tends to be stable as a whole with minor fluctuations.

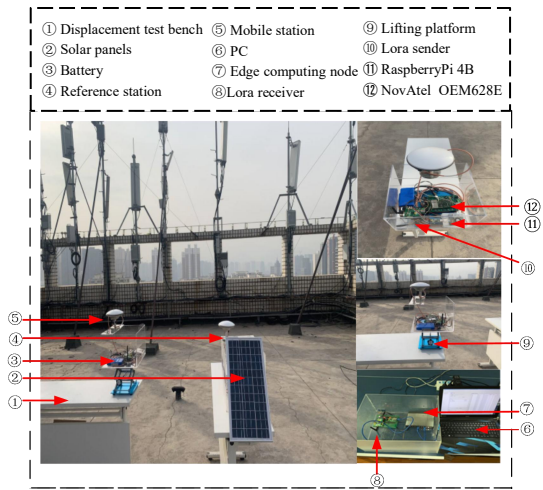

Figure 6: Monitoring System

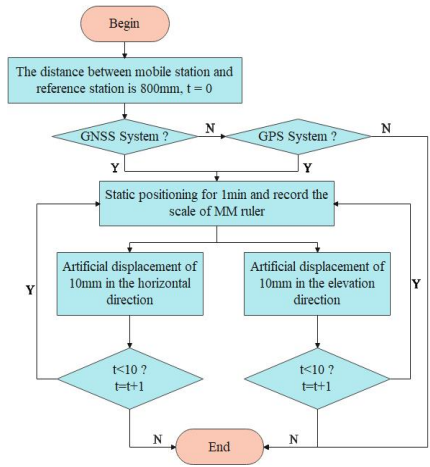

Figure 7: Displacement experiment flow chart

### 4.3 Result analysis of gross error elimination and data denoising

The GNSS combined system and the Single-GPS system are used for positioning, and artificial excitation sources are given to simulate landslide disasters in the horizontal and elevation directions. The specific steps are shown in Figure 7. Finally, the high-precision millimeter ruler measurement value is used as the actual value to compare and verify the original value. The system uses the improved Raida (3σ)criterion and Butterworth filter at the edge nodes to monitor the accuracy of real-time deformation after data processing.

The self-built deformation platform is used for comparative experiments, and The specific experimental results are shown in Figure 8. Through the analysis of the baseline monitoring results, it can be found that after the improved Raida criterion and low-pass filtering, the monitoring accuracy of the GNSS combined system is better than that of the GPS single system. The maximum horizontal error is improved from ±14.7mm to ±7.8mm, and the minimum error was improved from ±0.91mm to ± 0.39mm. The maximum error in the vertical direction was improved from ±10.2mm to ±8.4mm, and the minimum error was improved from ± 2.1mm to ± 0.1mm. The system uses GNSS-RTK technology to achieve millimeter-level deformation monitoring accuracy in horizontal and vertical directions.

## 5. CONCLUSION

This study developed a low-cost, high-precision landslide monitoring system using NovAtel's high-precision GNSS board and carrier phase differential technology (RTK). A "time slice" Raida criterion(3σ) was proposed to ensure that the timeliness of the error elimination is not the deformation data is eliminated as gross errors at the same time, and the calculation time is also shortened. In addition, the system also uses edge computing and Alibaba Cloud to achieve data processing and persistence and displays specific results on a website.

Due to the particularity of landslides and other geological disasters, this system comparison test is conducted in the form of artificially generated excitation sources. More experiments and tests are needed to study the efficiency of this system in real-time displacement monitoring in actual landslide scenarios.

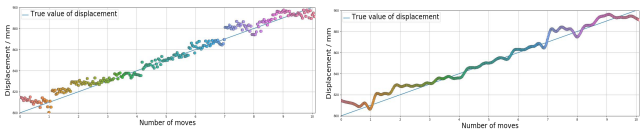

(a) GPS: horizontal gross error elimination and low-pass filtering

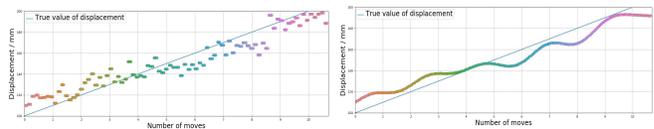

(b) GPS: height direction gross error elimination and low-pass filtering

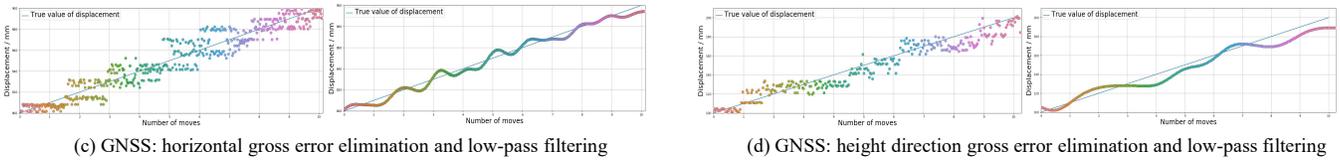

(c) GNSS: horizontal gross error elimination and low-pass filtering  (d) GNSS: height direction gross error elimination and low-pass filtering

Figure 8: Displacement experiment flow char